\newcommand{\be}[0]{\begin{equation}}
\newcommand{\ee}[0]{\end{equation}}
\newcommand{\ba}[0]{\begin{eqnarray}}
\newcommand{\ea}[0]{\end{eqnarray}}

\documentclass[epj,axodraw4j]{svjour}
\usepackage{graphics}
\usepackage{axodraw4j}
\usepackage{pstricks}
\usepackage{color}
\usepackage{amssymb}
\usepackage{epsfig}
\usepackage{graphics}
\usepackage{graphics}
\usepackage{axodraw4j}
\usepackage{pstricks}
\usepackage{color}
\usepackage{amssymb}
\usepackage{epsfig}
\authorrunning{H.~Nematollahi et al.}
\titlerunning{Unpolarized TMD densities based on the modified chiral quark model}
\begin{document}
\title{Unpolarized transverse momentum dependent densities based on the modified chiral quark model}
\author{H.~Nematollahi \inst{1}\and M.~M.~Yazdanpannah\inst{1}\thanks{\emph{myazdan@uk.ac.ir}}  \and A.~Mirjalili\inst{2}
%\thanks{\emph{}}
}                     % Do not remove

\offprints{} \institute{
 Faculty of physics, Shahid Bahonar University of Kerman, Kerman,
Iran  \and Physics Department, Yazd University, P.O.
Box 89195-741,  Yazd, Iran}
\date{Received: date / Revised version: date}
\abstract{ We investigate the transverse momentum dependent (TMD) quark and gluon
distribution functions in the modified chiral quark model ($\chi QM$).
Calculations of the TMD quark and gluon densities, using the modified $\chi QM$
are done for the first time in this article. For this propose we first
formulate the TMD interactions that occur in the $\chi QM$ at low $Q^2$ scale
($Q^2=0.35~ GeV^2$) and then obtain the TMD parton distributions inside the
proton, considering the interactions. To this end, we need to compute the TMD
bare quark distributions. These TMD bare densities are calculated, using the
solution of Dirac equation with a squared radial symmetry potential . It is
shown that our results consist appropriate behavior which are expecting for the
TMD parton distributions.  }  \maketitle
\section{Introduction}
To extend our knowledge of the nucleon structure far beyond what we know from parton distribution functions (PDFs) about longitudinal momentum distributions, we need a generalization of PDFs which are known as transverse momentum dependent parton distribution functions (TMDs). These functions contain also some information on transverse parton momenta as well as spin-orbit correlations.

TMDs are important since they play essential roles in the theoretical description of some experimental quantities like single spin asymmetries which exist in various hard processes including semi-inclusive deep inelastic scattering (SIDIS), Drell-Yan processes, etc \cite{SMC,HER,COM}. These functions can also be called the unintegrated parton distributions \cite{JCDS}.
\begin{figure}
\begin{center}
\begin{picture}{(400,50)(0,0)
\label{graph} \SetColor{Black} \SetScale{1}
 {\SetWidth{1.5}\Line(0,10)(30,10)
\Text(15,5)[t]{\small{$U$}}} \PText(35,18)(0)[t]{=}
   \Line(40,10)(70,10)
\Text(55,5)[t]{\small{$u_0$}} \PText(75,18)(0)[t]{+}
\Text(83,5)[t]{\small{$u_0$}} \Text(100,5)[t]{\small{$u$}}
\Text(117,5)[t]{\small{$u_0$}}
    \Line(80,10)(120,10)
\PText(125,18)(0)[t]{+}
    \DashCArc(100,10)(12,0,180){2}
\Text(105,35)[t]{\small{$\pi^{0}$}}
    \Vertex(88,10){1.5}
    \Vertex(112,10){1.5}
    \Line(130,10)(170,10)
    \DashCArc(150,10)(12,0,180){2}
    \Vertex(138,10){1.5}
    \Vertex(162,10){1.5}
\Text(155,35)[t]{\small{$\pi^{+}$}} \PText(175,18)(0)[t]{+}
\Text(133,5)[t]{\small{$u_0$}} \Text(151,6)[t]{\small{$d$}}
\Text(168,5)[t]{\small{$u_0$}}
    \Line(180,10)(220,10)
    \DashCArc(200,10)(12,0,180){2}
    \Vertex(188,10){1.5}
    \Vertex(212,10){1.5}
   \Line(180,10)(220,10)
    \DashCArc(200,10)(12,0,180){2}
    \Vertex(188,10){1.5}
    \Vertex(212,10){1.5}
\Text(205,35)[t]{\small{$K^{+}$}} \PText(225,18)(0)[t]{+}
    \Line(230,10)(270,10)
\Text(183,5)[t]{\small{$u_0$}} \Text(201,6)[t]{\small{$s$}}
\Text(218,5)[t]{\small{$u_0$}}
    \GlueArc(250,10)(12,0,180){2}{8}
\Text(250,35)[t]{\small{$g$}}
   \Vertex(238,10){1.5}
    \Vertex(262,10){1.5}
\Text(233,5)[t]{\small{$u_0$}} \Text(250,6)[t]{\small{$u$}}
\Text(268,5)[t]{\small{$u_0$}} \PText(280,18)(0)[t]{+ ...}
\SetColor{Black}
%%%%%%%%%%%%%%%%%%%%%%%%%%%%%%%%%
\SetColor{Black} \SetScale{1}{\SetWidth{1.5}
 \Line(0,-45)(30,-45)
\Text(15,-50)[t]{\small{$D$}}} \PText(35,-37)(0)[t]{=}
   \Line(40,-45)(70,-45)
\Text(55,-50)[t]{\small{$d_0$}} \PText(75,-37)(0)[t]{+}
\Text(83,-50)[t]{\small{$d_0$}} \Text(100,-50)[t]{\small{$d$}}
\Text(117,-50)[t]{\small{$d_0$}}
    \Line(80,-45)(120,-45)
\PText(125,-37)(0)[t]{+}
    \DashCArc(100,-45)(12,0,180){2}
\Text(105,-20)[t]{\small{$\pi^{0}$}}
    \Vertex(88,-45){1.5}
    \Vertex(112,-45){1.5}
    \Line(130,-45)(170,-45)
    \DashCArc(150,-45)(12,0,180){2}
    \Vertex(138,-45){1.5}
    \Vertex(162,-45){1.5}
\Text(155,-20)[t]{\small{$\pi^{-}$}} \PText(175,-37)(0)[t]{+}
\Text(133,-50)[t]{\small{$d_0$}} \Text(151,-49)[t]{\small{$u$}}
\Text(168,-50)[t]{\small{$d_0$}}
   \Line(180,-45)(220,-45)
    \DashCArc(200,-45)(12,0,180){2}
    \Vertex(188,-45){1.5}
    \Vertex(212,-45){1.5}
\Text(205,-20)[t]{\small{$K^{0}$}} \PText(225,-37)(0)[t]{+}
    \Line(230,-45)(270,-45)
\Text(183,-50)[t]{\small{$d_0$}} \Text(201,-49)[t]{\small{$s$}}
\Text(218,-50)[t]{\small{$d_0$}}
    \GlueArc(250,-45)(12,0,180){2}{8}
\Text(250,-20)[t]{\small{$g$}}
   \Vertex(238,-45){1.5}
    \Vertex(262,-45){1.5}
\Text(233,-50)[t]{\small{$d_0$}} \Text(250,-49)[t]{\small{$d$}}
\Text(268,-50)[t]{\small{$d_0$}} \PText(280,-37)(0)[t]{+ ...}
\SetColor{Black}}
\end{picture}
\vspace{2cm}
\caption{\small Up ($U$) and down ($D$) constituent quarks. This figure is quoted from Ref.\cite{nymf}. }\label{fluctuation} \label{fig:1}
\end{center}
\end{figure}
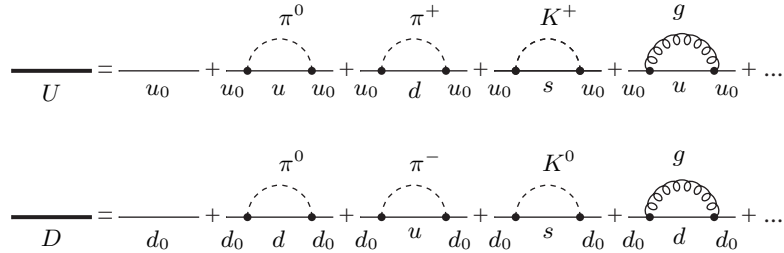

TMDs are expecting to perform us a three-dimensional view of the parton distributions in momentum space. Therefore they give us sufficient information in addition to what can be learned from the generalized parton densities \cite{MB,RP,MD,BJY}.

In comparison with the integrated parton densities, our knowledge about the TMD parton distributions is not rich. In contrast to
the integrated parton distributions which there are many model calculations for them, there are not enough for the TMD distributions.

The purpose of this work is extracting the unpolarized TMD quark and gluon distributions using the chiral quark model.\\
The main property of the $\chi QM$ is its application at low $Q^2$ scales. This property is due to breaking the chiral symmetry at low energy scales. In these scales, the light quark masses are not ignorable in comparison with the nucleon mass and chiral symmetry is broken.\\
In the $\chi QM$, the bare quarks inside the nucleon are surrounded by the clouds of Goldstone (GS) bosons and gluons \cite{myj,nymf}. Considering the interactions which occur in this bounded system at the first approximation and also the relations between the transverse momentum of the bare quarks and GS bosons(gluons) in these interactions, we can obtain the TMD quark and gluon distributions inside the proton. In order to do these calculations, we need to know the proton TMD bare quark distributions. Using the solution of the  Dirac equation by considering the harmonic oscillator potential, these TMD bare distributions can be calculated.
\begin{figure*}
\begin{center}
\fcolorbox{white}{white}{
  \begin{picture}(451,71) (231,-77)
    \SetWidth{1.0}
    \SetColor{Black}
    \Line(232.128,-56.505)(351.247,-56.505)
    \Line[dash,dashsize=1.527](281.761,-56.505)(310.777,-27.489)
    \Text(232.128,-65.668)[lb]{\Large{\Black{$$}}}
    \Text(341.32,-65.668)[lb]{\Large{\Black{$$}}}
    \Text(284.761,-36.652)[lb]{{\Black{${\cal B}$}}}
    \Text(354.301,-80.176)[lb]{{\Black{$$}}}
    \Text(345.902,-69.486)[lb]{\Large{\Black{$$}}}
    \Text(361.716,-81.703)[lb]{\large{\Black{$(a)$}}}
    \Line(610.864,-58.032)(500.908,-58.032)
    \Gluon(546.723,-58.032)(574.212,-30.543){2.644}{4}
    \Text(548.723,-38.943)[lb]{{\Black{$g$}}}
    \Text(500.908,-66.431)[lb]{\Large{\Black{$$}}}
    \Text(601.701,-66.431)[lb]{\Large{\Black{$$}}}
    \Text(616.209,-80.176)[lb]{\Large{\Black{$$}}}
    \Text(610.864,-69.486)[lb]{\Large{\Black{$$}}}
    \Text(630.915,-81.703)[lb]{\large{\Black{$(b)$}}}
    \Text(236.71,-70.939)[lb]{{\Black{$q_i$}}}
    \Text(346.665,-70.939)[lb]{{\Black{$q_j$}}}
    \Text(506.253,-71.703)[lb]{{\Black{$q_i$}}}
    \Text(610.1,-71.176)[lb]{{\Black{$q_j$}}}
  \end{picture}
} \caption{\small (a)  The fluctuation of a bare quark $q_{i}$ into
a GS boson ${\cal B}$ plus a struck quark $q_{j}$. (b) A bare quark $q_{i}$ emits a gluon and transforms to a constituent quark $q_{j}$.}\label{fluctuation} \label{fig:2}
\end{center}
\end{figure*}
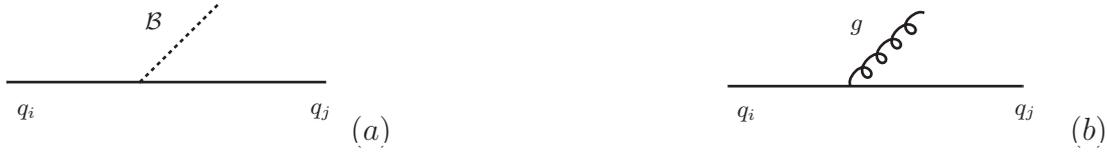

The plan of this paper is as follows. Applying the $\chi QM$, the TMD quark and gluon densities are calculated in section 2. For this purpose, we first investigate the interactions which occur at the vertex of bare quark-GS bosons and also the bare quark-gluon vertex in the $\chi QM$ in subsections 2.1 and 2.2. In subsection 2.3 the required TMD bare quark distributions are computed. We give our results in section 3 and finally render our conclusions in section 4.
\section{TMD quark and gluon distributions in the chiral quark model}
In this section, we calculate the transverse momentum dependence  of unpolarized quark and gluon distribution functions applying the chiral quak model. In this model which is used at low $Q^2$ scales, the important degrees of freedom are expressed in terms of quarks, gluons and GS bosons. In the $\chi QM$, the nucleon is consisted of the bare up and down quarks $(u_0, d_0)$ which are surrounded by the clouds of GS bosons and gluons (Fig.1). We first investigate in subsection 2.1 and 2.2 the different types of interactions which can be occurred at the vertexes of Fig.1.
\subsection{The bare quark-GS boson vertex}
\begin{figure*}[htp]
  \begin{center}
    \begin{tabular}{ccc}
      {\includegraphics[width=50mm,height=50mm]{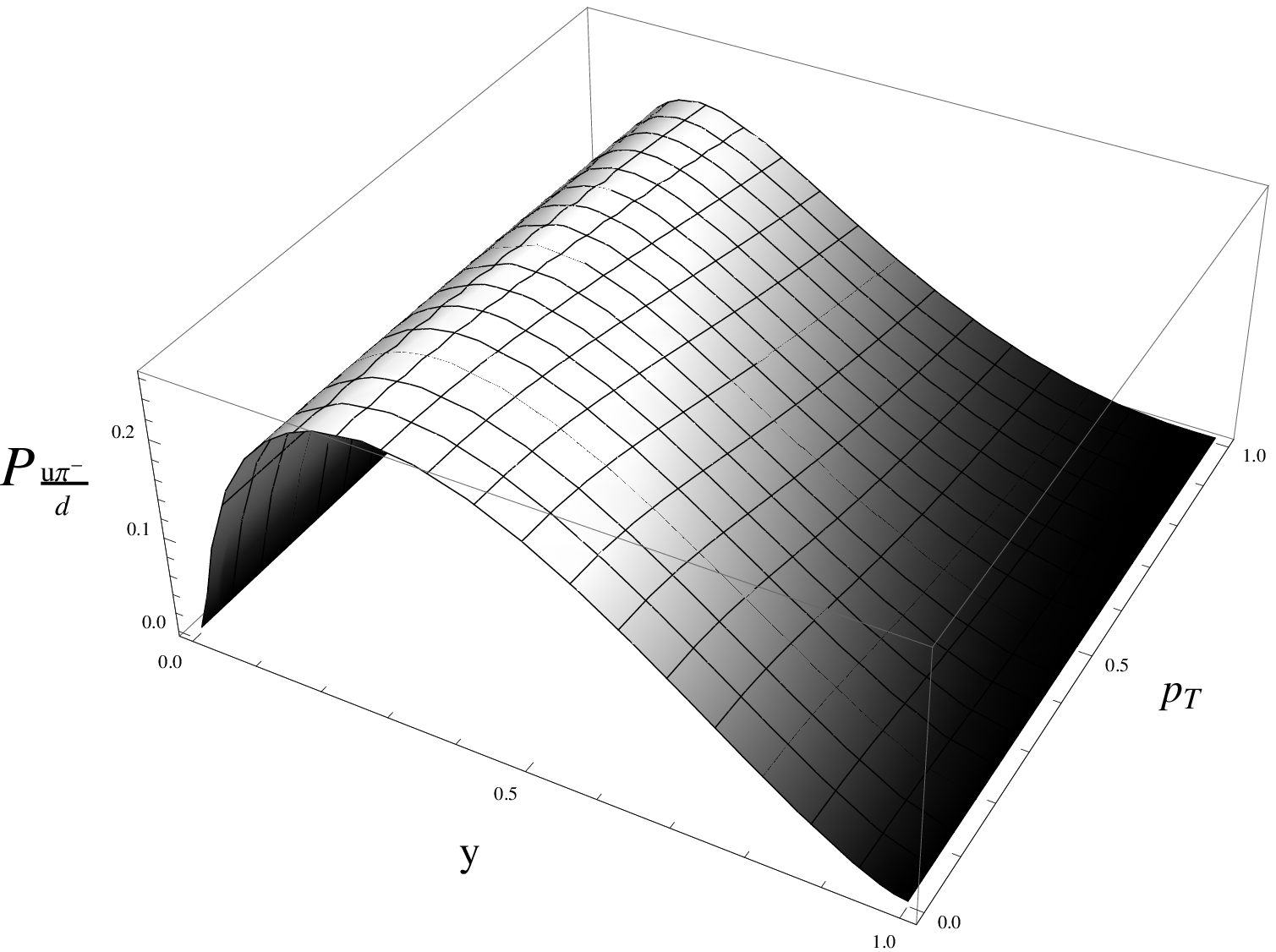}} &
      {\includegraphics[width=50mm,height=50mm]{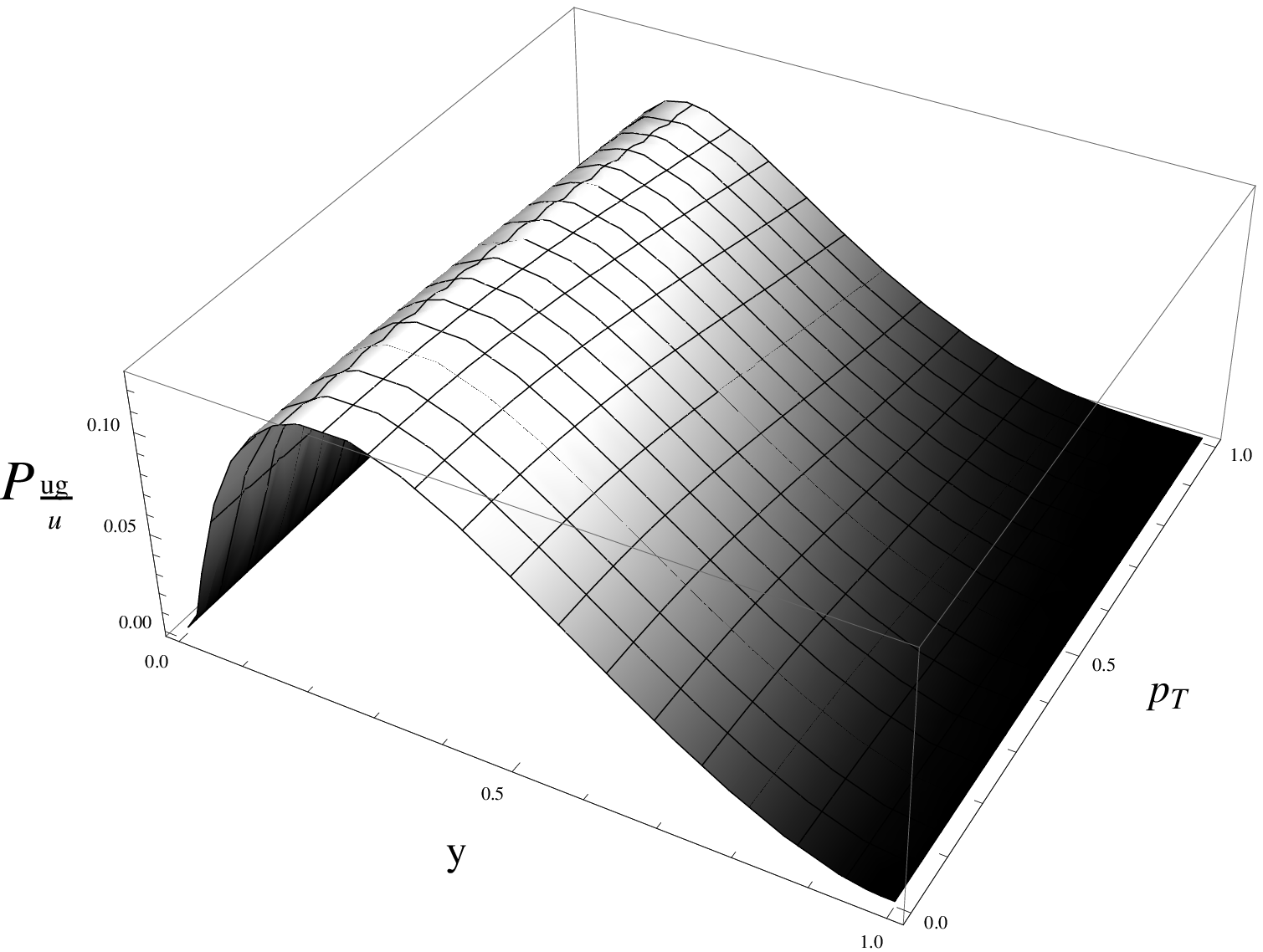}} &
       {\includegraphics[width=50mm,height=50mm]{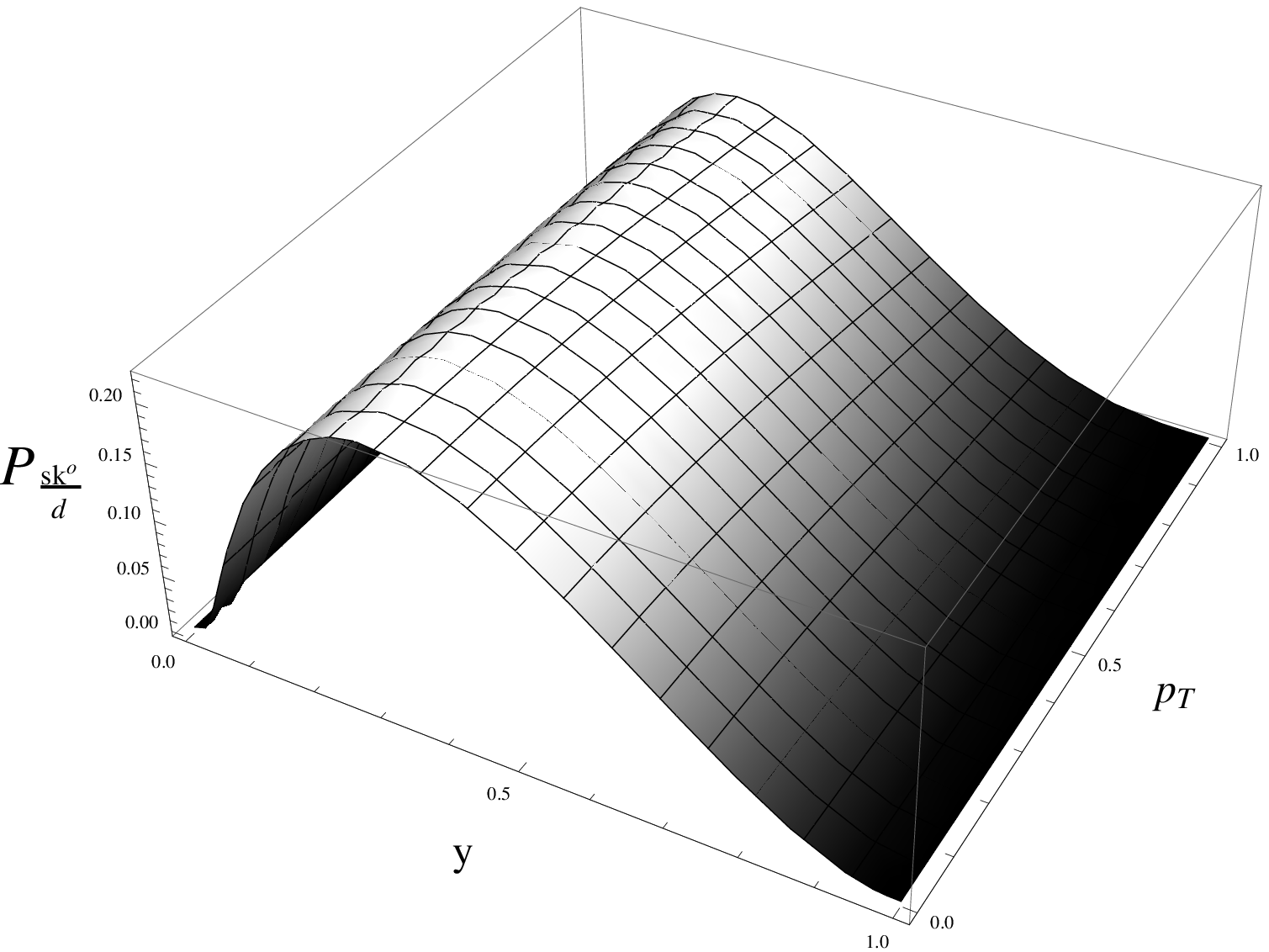}} \\
     \end{tabular}
\caption{ The three dimentional representaions of  transverse momentum dependent splitting functions $P_{u\pi^{-}/d}$, $P_{{ug}/{u}}$ and $P_{sk^{0}/d}$ with respect to $y$ and $p_T$. \label{fig:3}}
      \end{center}
\end{figure*}
At the first approximation, one basic process which can be occurred at the vertex of bare quark-GS boson in Fig.1 is considered. In this process, the bare quark $q_i$ fluctuates into the intermediate quark $q_j$ and a GS boson ${\cal B}$ which is represented in Fig.2(a). This fluctuation can be given by:
\begin{equation}
q_{j}(x,p_{T})=\int^{1}_{x}\frac{\textmd{d}y}{y}P_{j{\cal
B}/i}(y,p_{T})q_{0,i}(\frac{x}{y},p_{T})\;.\label{qq}
\end{equation}
In Eq.(\ref{qq}), $P_{j{\cal B}/i}$ is the splitting function that expresses the probability for finding the struck quark $q_j$ with longitudinal momentum fraction $y$ and the transverse momentum ${p_{j}}_{T}$ and also the GS boson which is carried the longitudinal momentum fraction $1-y$ of the parent quark's momentum and the transverse momentum ${p_{\cal B}}_{T}$.\\
We can write the transverse momenta of the quark $q_j$ and the GS boson in terms of the transverse momentum of the bare quark $q_i$, $p_{T}$, and the intrinsic variables \cite{pasquini}:
\begin{eqnarray}\label{intrinsic}
{p_{j}}_{T}=k_{T}+yp_{T},~{p_{\cal B}}_{T}=-k_{T}+(1-y)p_{T}
 \;.
\end{eqnarray}
Based on Sullivan processes \cite{Suli} we can suggest the TMD splitting function $P_{j{\cal B}/i}$ as it follows:
\begin{eqnarray}\label{qsplit}
P_{j{\cal B}/i}(y,p_{T})=\frac{1}{8\pi^{2}}(\frac{g_{A}\bar{m}}{f})^{2}(1-y)\nonumber\\
\int^{t_{min}}_{-\Lambda^2_{\chi}} \frac{[(m_{i}-m_{j})^{2}-t]}{(t-m^{2}_{{\cal
B}})^{2}}\textmd{d}t\;,
\end{eqnarray}
where $\Lambda_{\chi}$ is the cut off parameter, $m_i$, $m_{j}$ and $m_{\cal B}$ are the mass of quarks $q_i$, $q_j$ and the GS boson, respectively. The parameter $t$ is defined as:
\begin{eqnarray}\label{t}
t=\frac{-[{p^2_{j}}_{T}+(1-y)[m^2_{j}-ym^2_{i}]]}{y}\nonumber\\
=\frac{-[(k_{T}+yp_{T})^2+(1-y)[m^2_{j}-ym^2_{i}]]}{y}\;.
\end{eqnarray}
Substituting the t parameter in Eq.(\ref{qsplit}), we will arrive at:
\begin{eqnarray}\label{qsplitting}
P_{j{\cal B}/i}(y,p_{T})=\int \textmd{d}k_{T}\frac{2(k_{T}+yp_{T})}{y^{2}(1-y)(m_{i}^{2}-M^{2}_{j{\cal B}})^{2}}\nonumber\\
((m_{j}-m_{i}y)^{2}+(k_{T}+yp_{T})^{2})\times \frac{1}{8\pi^{2}}(\frac{g_{A}\bar{m}}{f})^{2}.
\end{eqnarray}
In Eq.(\ref{qsplitting}), $g_{A}$ and $f$ denote the axial vector constant and the pseudo-scalar decay constant, respectively; $\bar{m}$ is the average mass of $q_i$ and $q_j$ and $M^{2}_{j{\cal B}}$ is the square invariant mass of the final stat:
\begin{equation}
M^{2}_{j{\cal B}}=\frac{m^2_{j}+(k_{T}+yp_{T})^2}{y}
+\frac{m^2_{{\cal B}}+(k_{T}+yp_{T})^2}{1-y}.
\end{equation}
It is obvious that $t_{min}$ in Eq.(\ref{qsplit}) can be obtained from Eq.(\ref{t}) by substituting $k_{T}=0$ in this equation \cite{EHQ}.
It is found that if we put $p_{T}=0$ in Eq.(\ref{t}), this equation will be casted to its usual form while there is not finally any transverse momentum dependence in the $\chi QM$  \cite{weber,weber1}. In this case the splitting function $P_{j{\cal B}/i}$ and also the $q_j$ distribution are given by expressions which are used in Refs.\cite{nymf,swnpa,MA2005,MA2011}.
\subsection{The vertex of the bare quark-gluon}
A similar process can be occurred in the vertex of bare quark-gluon, at the first approximation (Fig.2(b)). In this process the bare quark $q_i$ emits a gluon and appears as the recoiled quark $q_j$. The TMD $q_j$ distribution has the following form:
\begin{equation}
q_{j}(x,p_{T})=\int^{1}_{x}\frac{\textmd{d}y}{y}P_{jg/i}(y,p_{T})q_{0,i}(\frac{x}{y},p_{T})\;.\label{qg}
\end{equation}
Here the related TMD splitting function $P_{jg/i}$ is:
\begin{eqnarray}\label{gsplitting}
P_{jg/i}(y,p_{T})=\int \textmd{d}k_{T} G^2_{jg/i}\frac{2(k_{T}+yp_{T})}{y^{2}(1-y)(m_{i}^{2}-M^{2}_{jg})^{2}}\nonumber\\
((m_{j}-m_{i}y)^{2}+(k_{T}+yp_{T})^{2})\times C_f\frac{{\alpha}_s(Q^2)}{4\pi}\;,
\end{eqnarray}
where $C_f$ is the color factor and the square invariant mass of the final quark-gluon state, $M^{2}_{jg}$, is written as:
\begin{equation}
M^{2}_{jg}=\frac{m^2_{j}+(k_{T}+yp_{T})^2}{y}
+\frac{m^2_{g}+(k_{T}+yp_{T})^2}{1-y}\;.
\end{equation}
In this equation $m_g$ denotes the gluon mass.\\
In Eq.(\ref{gsplitting}), the vertex function $G_{jg/i}$ is defined as:
\begin{equation}
G_{jg/i}= \exp \left ( {m_i^2-M_{jg}^2
\over 2\Lambda_{\chi} ^2} \right ).
\end{equation}
The TMD gluon distribution is also calculated via the interaction of Fig.2(b).

We use the following notation for the convolution integrals in Eqs.(\ref{qq}) and (\ref{qg}):
\begin{eqnarray}
P_{j{\cal B}/i}\otimes q_0=\int^{1}_{x}\frac{\textmd{d}y}{y}P_{j{\cal
B}/i}(y,p_{T})q_{0}(\frac{x}{y},p_{T}),\nonumber\\
P_{jg/i}\otimes q_0=\int^{1}_{x}\frac{\textmd{d}y}{y}P_{jg/i}(y,p_{T})q_{0}(\frac{x}{y},p_{T}).
\end{eqnarray}
\subsection{TMD bare quark distribution functions}
In order to calculate the TMD distribution functions inside the proton using Eqs.(\ref{qq}) and (\ref{qg}), we should first compute the TMD bare quark distributions. For this purpose we use the solution of  the Dirac equation under harmonic oscillator potential \cite{nymf,phdt,ydm}. Applying this approach the ground state wave function of the bare quark in the momentum space in terms of two parameters $\rho$ and $R$ is obtained as \cite{nymf}:
\begin{equation}
\phi_{0}(p)=-\pi^{-\frac{3}{4}}R^{\frac{3}{2}}(1+\frac{3\rho^{2}}{2})^{-\frac{1}{2}}
e^{-\frac{p^{2}R^{2}}{2}}\chi_{s}\chi_{f}\chi_{c}\;,
\end{equation}
where $\chi_{s}$, $\chi_{f}$ and $\chi_{c}$ are the related spin, flavor and color parts of the wave function.\\
We consider the probability density as $\varrho=\phi^\dagger_{0}(p)\phi_{0}(p)$.\\
The TMD bare quark distribution, $f(x,p_{T})$, satisfies the following relation \cite{nymf,MA2009}:
\begin{eqnarray}
\int\varrho\delta(p^{0}-\sqrt{(p^{3})^{2}+({p}_{T})^{2}+m^{2})}\textmd{d}p^{0}\textmd{d}p^{3}\textmd{d}^{2}p_{T}\nonumber\\
=\int f(x,{p}_{T})\textmd{d}^{2}p_{T}\textmd{d}x\;.
\label{integer}
\end{eqnarray}
In above equation $p^0$, $\vec{p}=(p^1,p^2,p^3)$ and $m$ are the bare quark energy, 3-momentum and mass, respectively. In Eq.(\ref{integer}) we use the relations which exist between the components of four vector momentum in standard coordinates and light cone coordinates so we can write \cite{nymf,MA2009}:
\begin{equation}
\textmd{d}p^{0}\textmd{d}p^{3}\textmd{d}^{2}p_{T}=\frac{1}{2}M_{t}\textmd{d}p^{-}\textmd{d}x\textmd{d}^{2}p_{T}\;.
\label{measure}
\end{equation}
\begin{figure*}[htp]
  \begin{center}
    \begin{tabular}{cc}
      {\includegraphics[width=60mm,height=40mm]{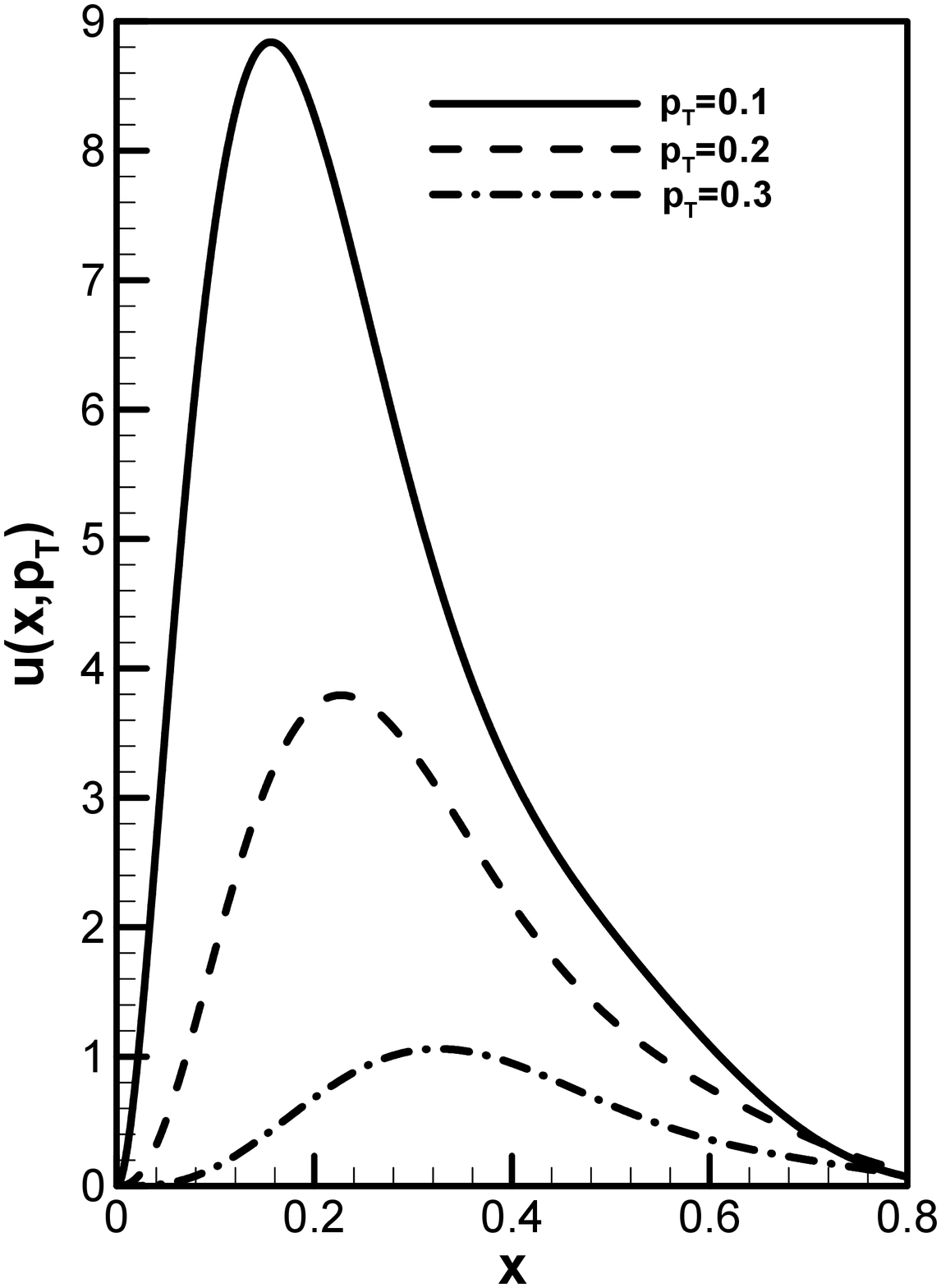}} &
       {\includegraphics[width=60mm,height=40mm]{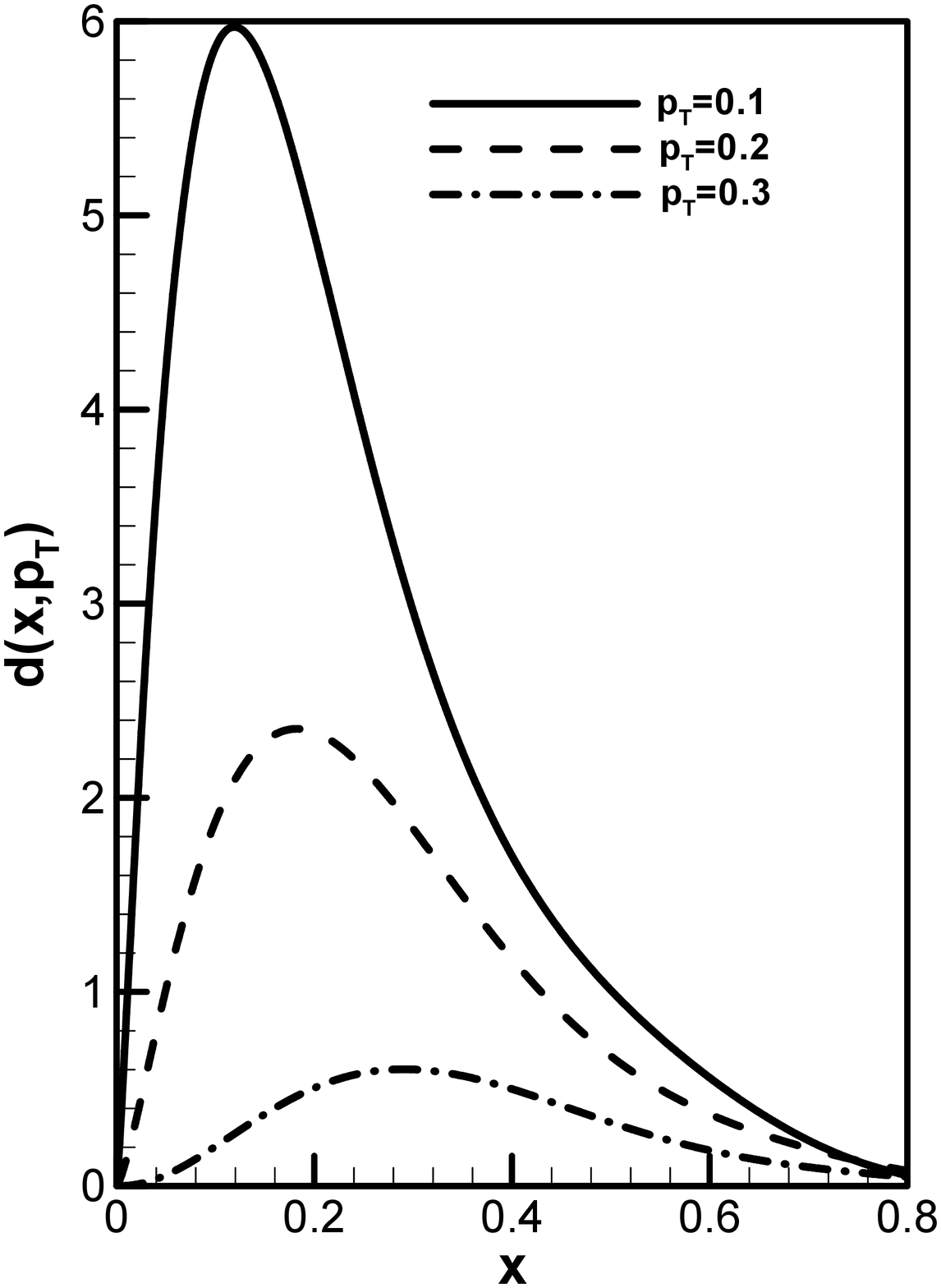}} \\
       {\includegraphics[width=60mm,height=40mm]{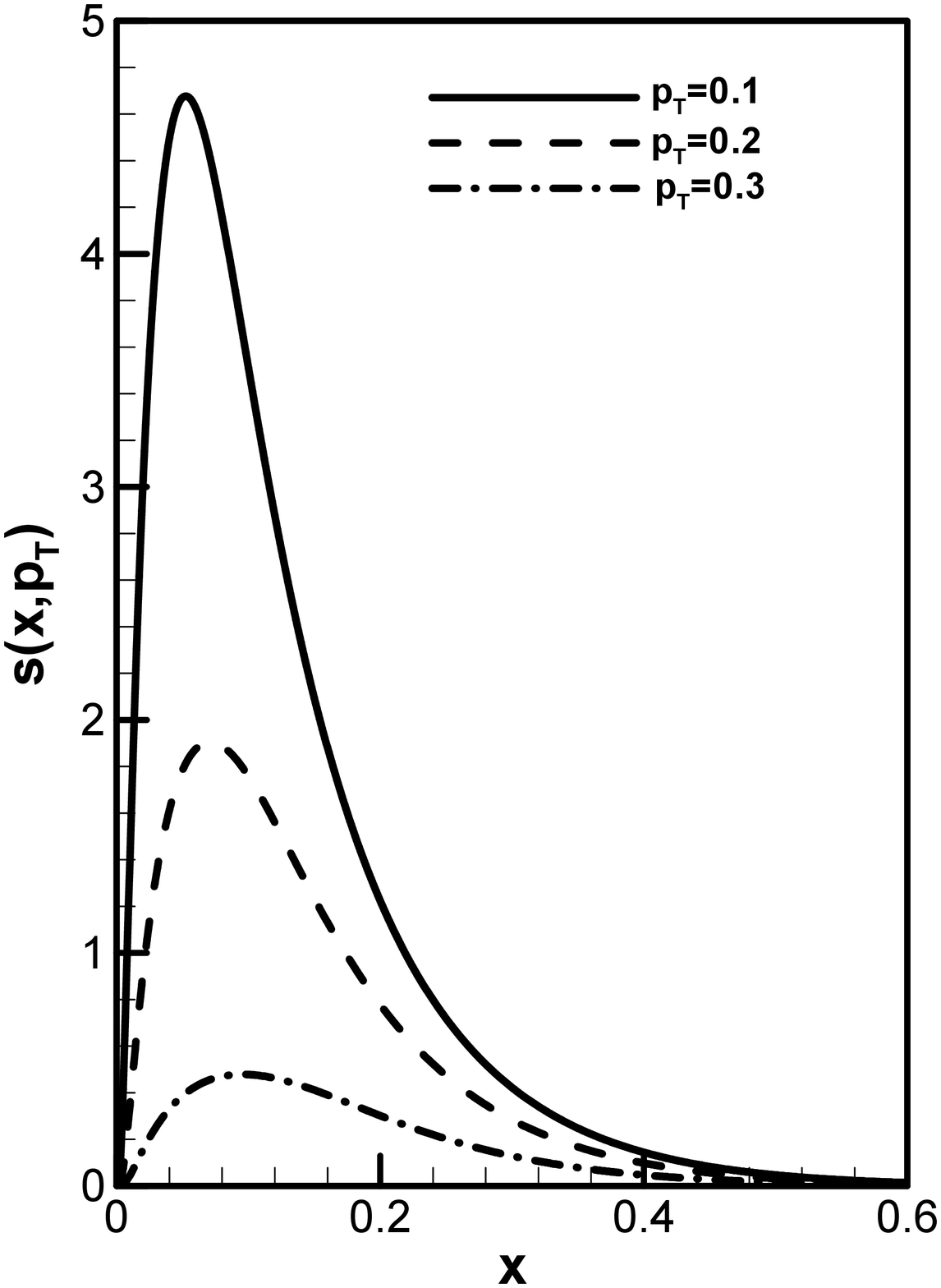}} &
       {\includegraphics[width=60mm,height=40mm]{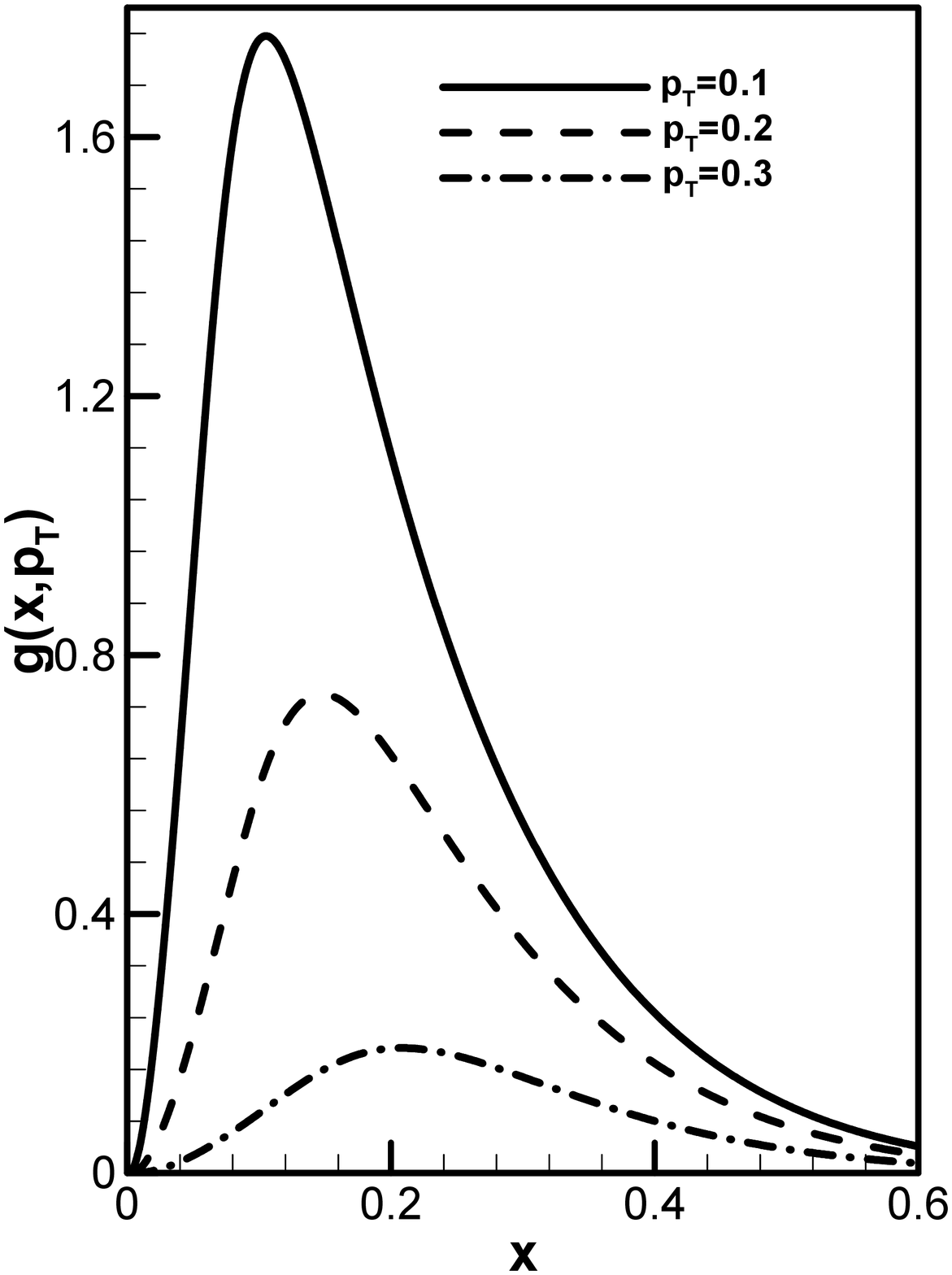}} \\
     \end{tabular}
\caption{ The TMD quark and gluon distribution functions with respect to $x$ at $p_T=0.1, 0.2$ and $0.3~GeV$. \label{fig:4}}
      \end{center}
\end{figure*}
Finally, by comparing the both sides of Eq.(\ref{integer}), the TMD bare quark distribution is determined as \cite{nymf}:
\begin{eqnarray}\label{distrib}
f(x,{p}_{T})=\frac{1}{2}M_{t}R^{3}\pi^{-\frac{3}{2}}(1+\frac{3\rho^{2}}{2})^{-1}[1+\frac{({p}_{T})^{2}+m^{2}}{(M_{t}x)^{2}}]\nonumber\\
\times e^{-R^{2}({p}_{T})^{2}}e^{-\frac{R^{2}}{4}[M_{t}x-\frac{({p}_{T})^{2}+m^{2}}{M_{t}x}]^{2}};f=u_0,d_0.\nonumber\\
\end{eqnarray}
\begin{figure*}[htp]
  \begin{center}
    \begin{tabular}{cc}
      {\includegraphics[width=60mm,height=40mm]{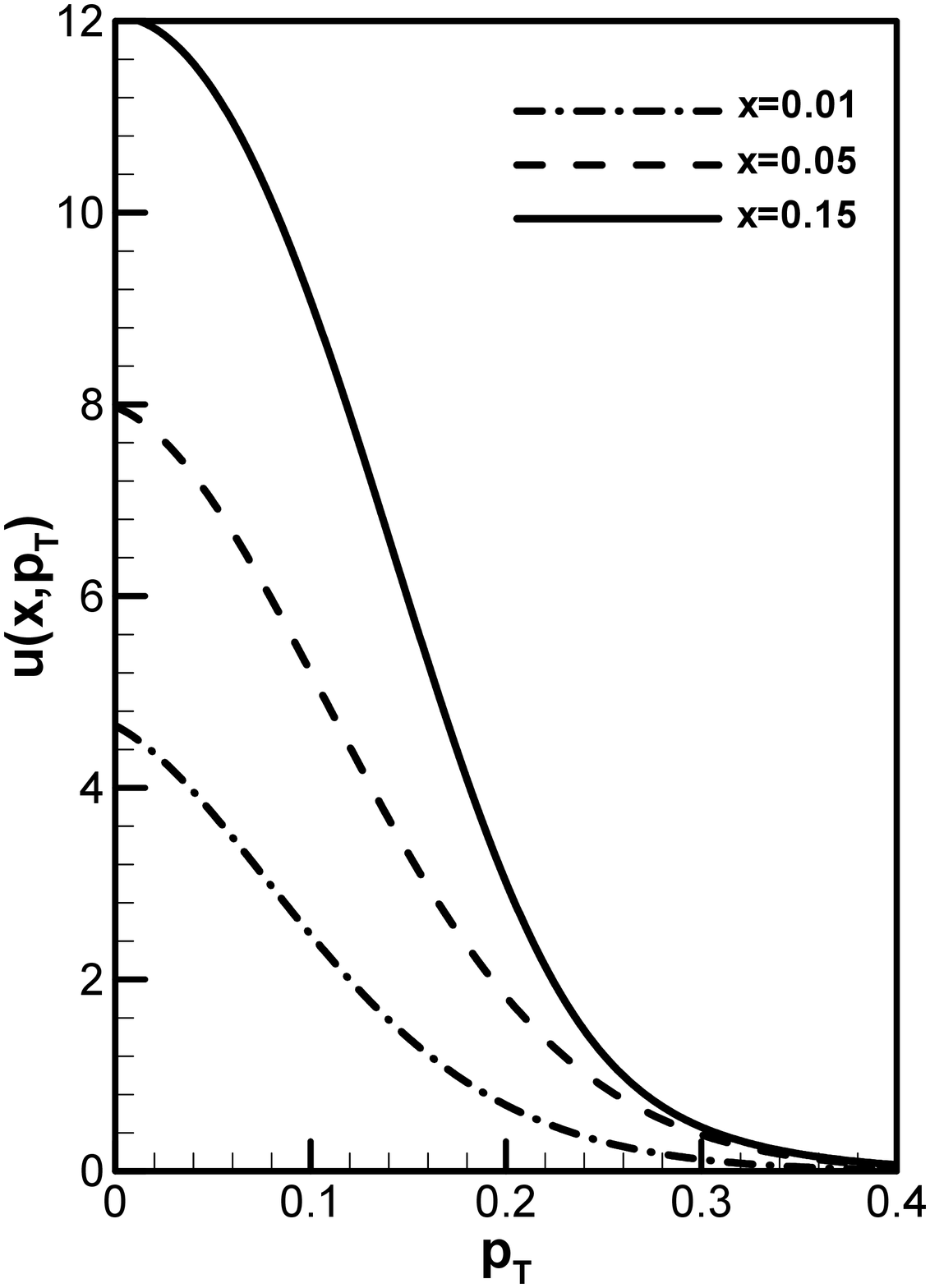}} &
       {\includegraphics[width=60mm,height=40mm]{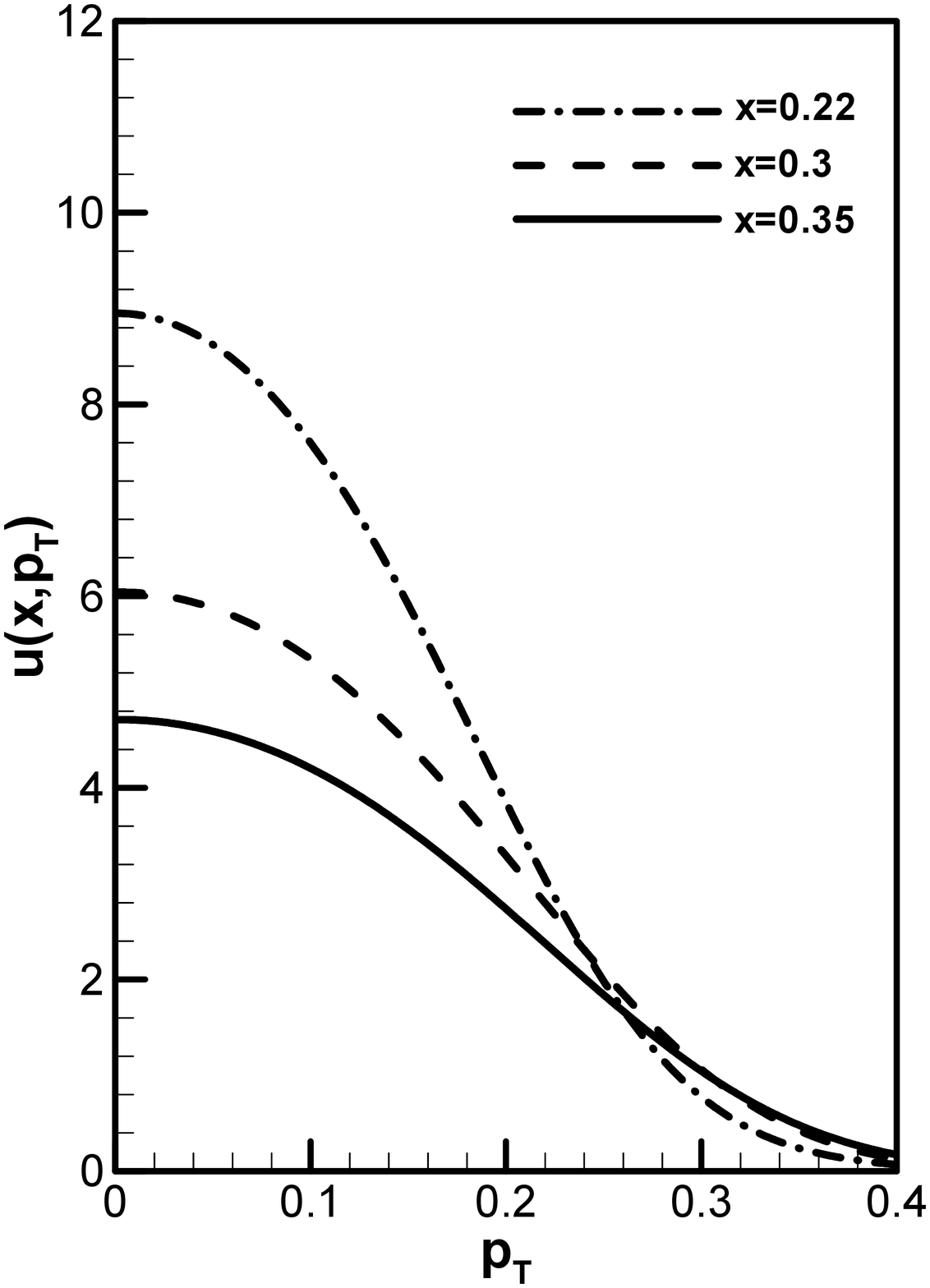}} \\
       {\includegraphics[width=60mm,height=40mm]{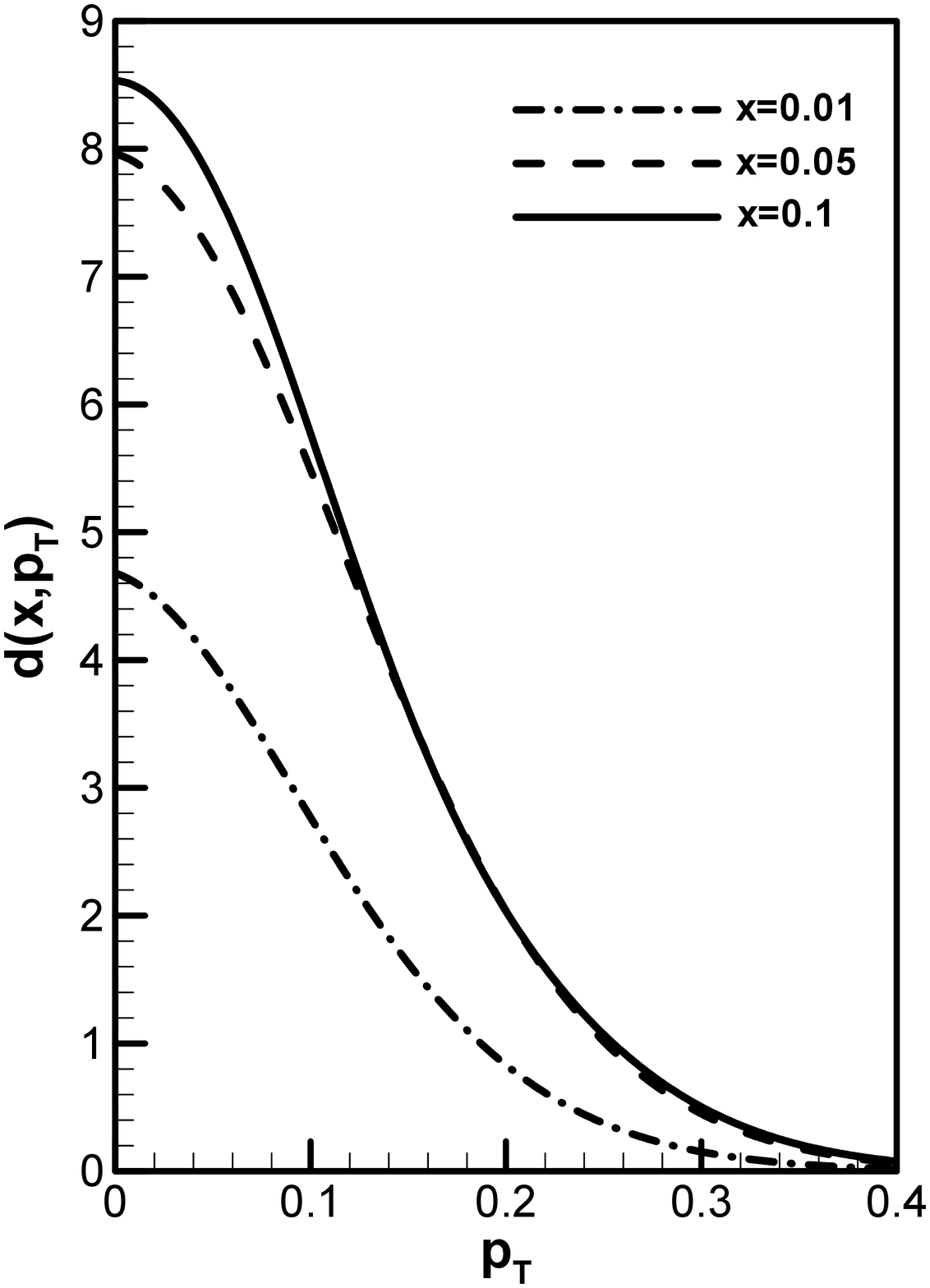}} &
       {\includegraphics[width=60mm,height=40mm]{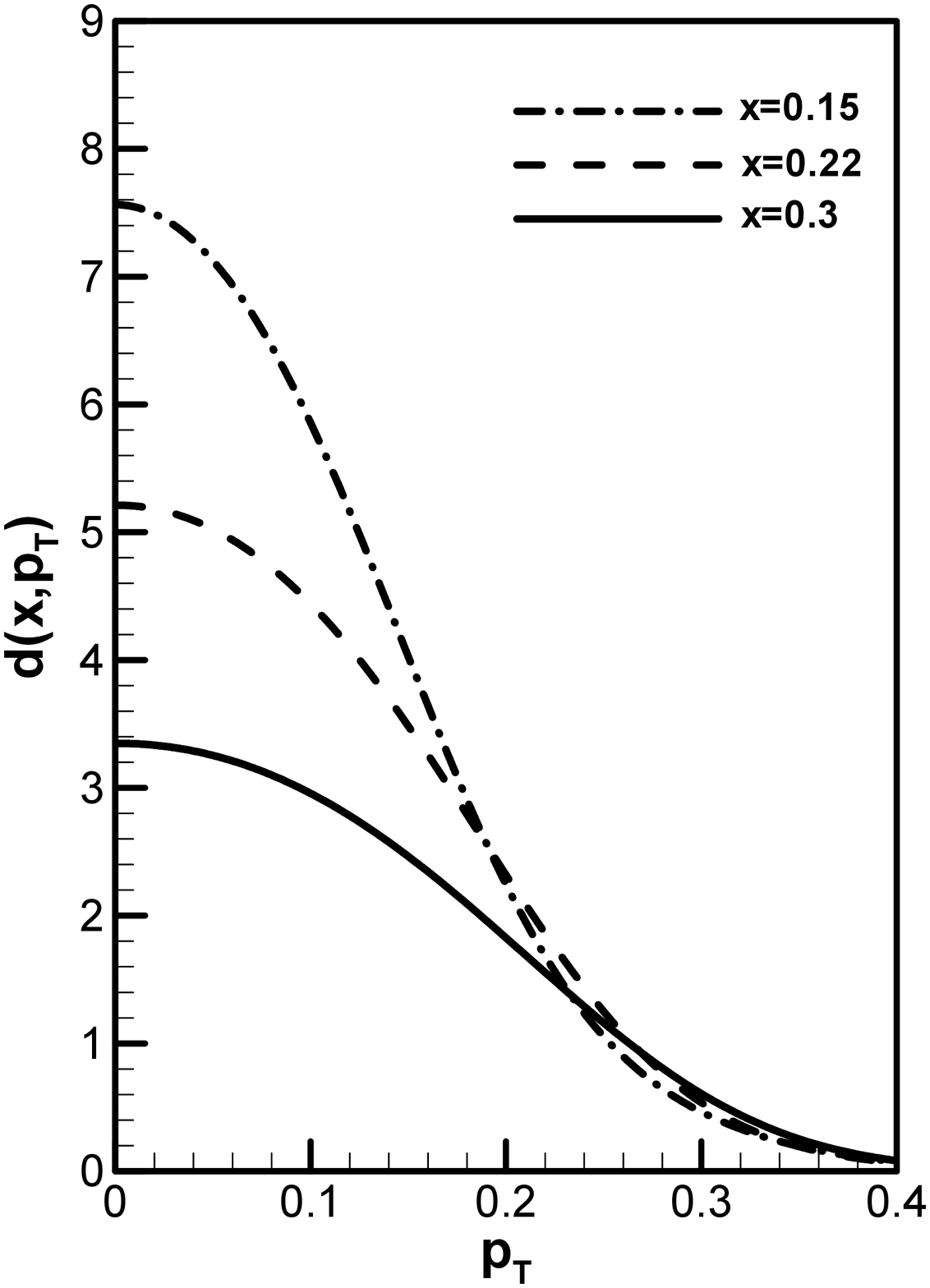}} \\
     \end{tabular}
\caption{ The TMD distribution of $u$ and $d$ quarks at two set of $x$ values. \label{fig:5}}
      \end{center}
\end{figure*}
\begin{figure*}[htp]
  \begin{center}
    \begin{tabular}{cc}
      {\includegraphics[width=60mm,height=40mm]{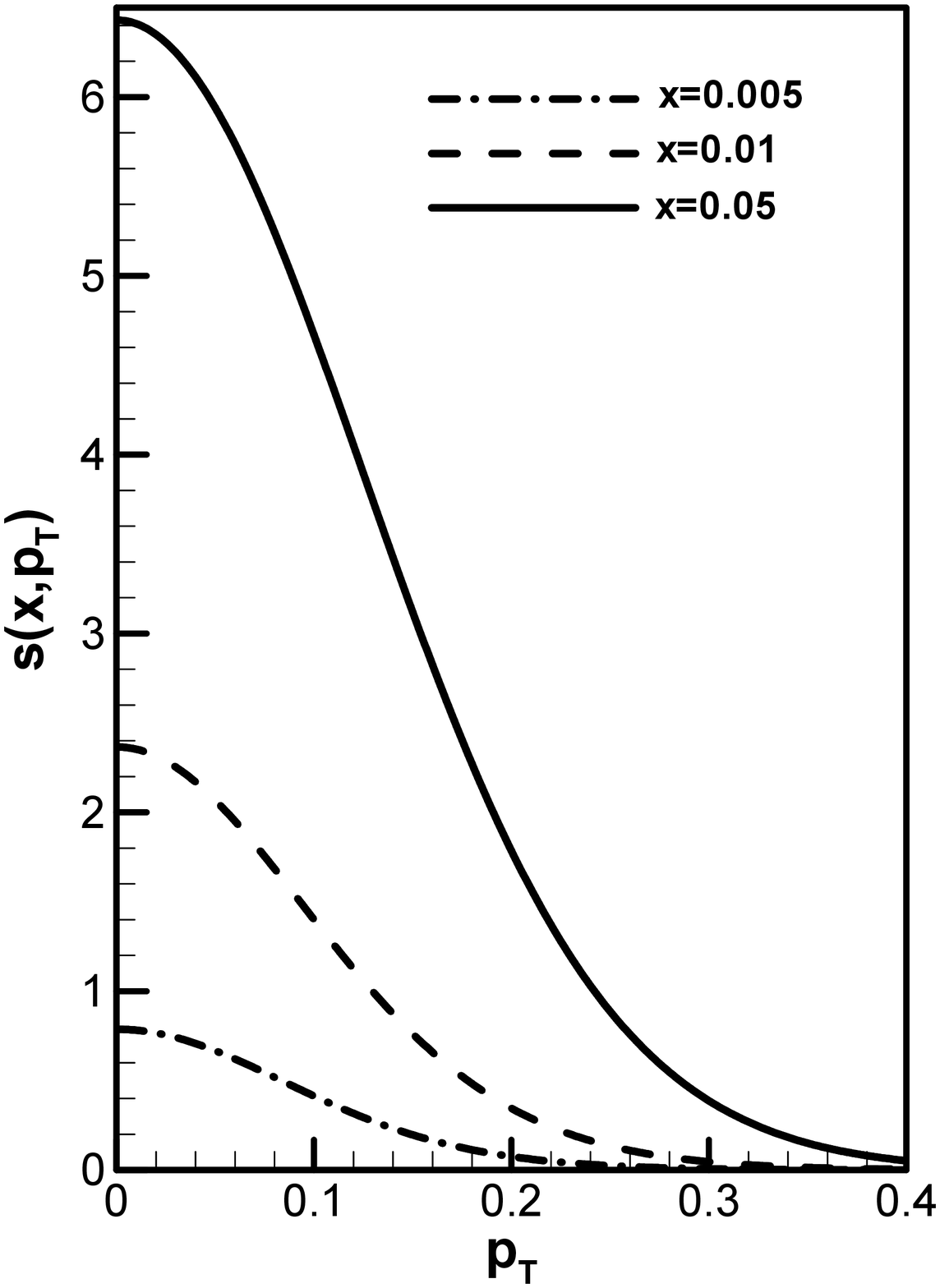}} &
       {\includegraphics[width=60mm,height=40mm]{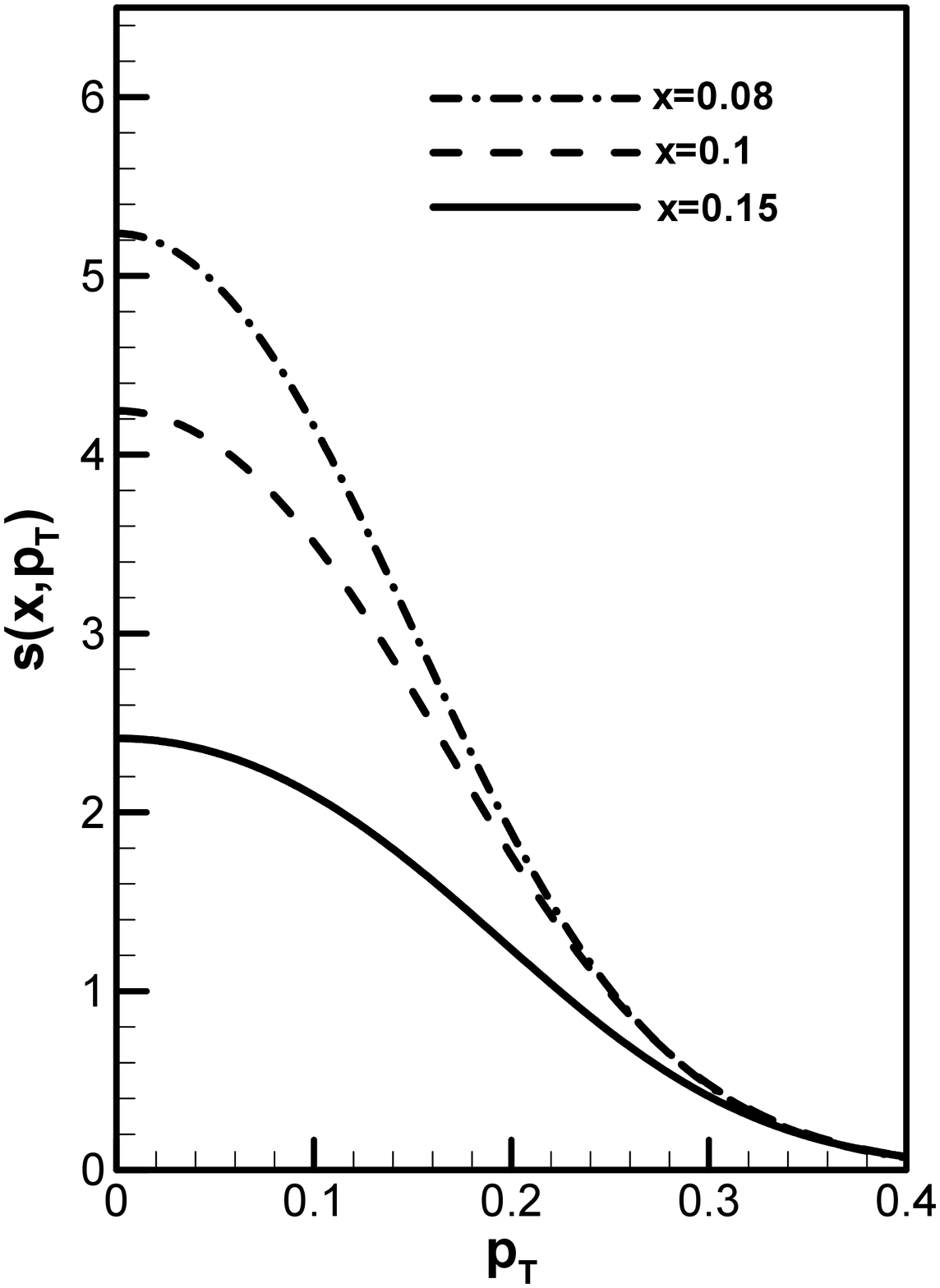}} \\
       {\includegraphics[width=60mm,height=40mm]{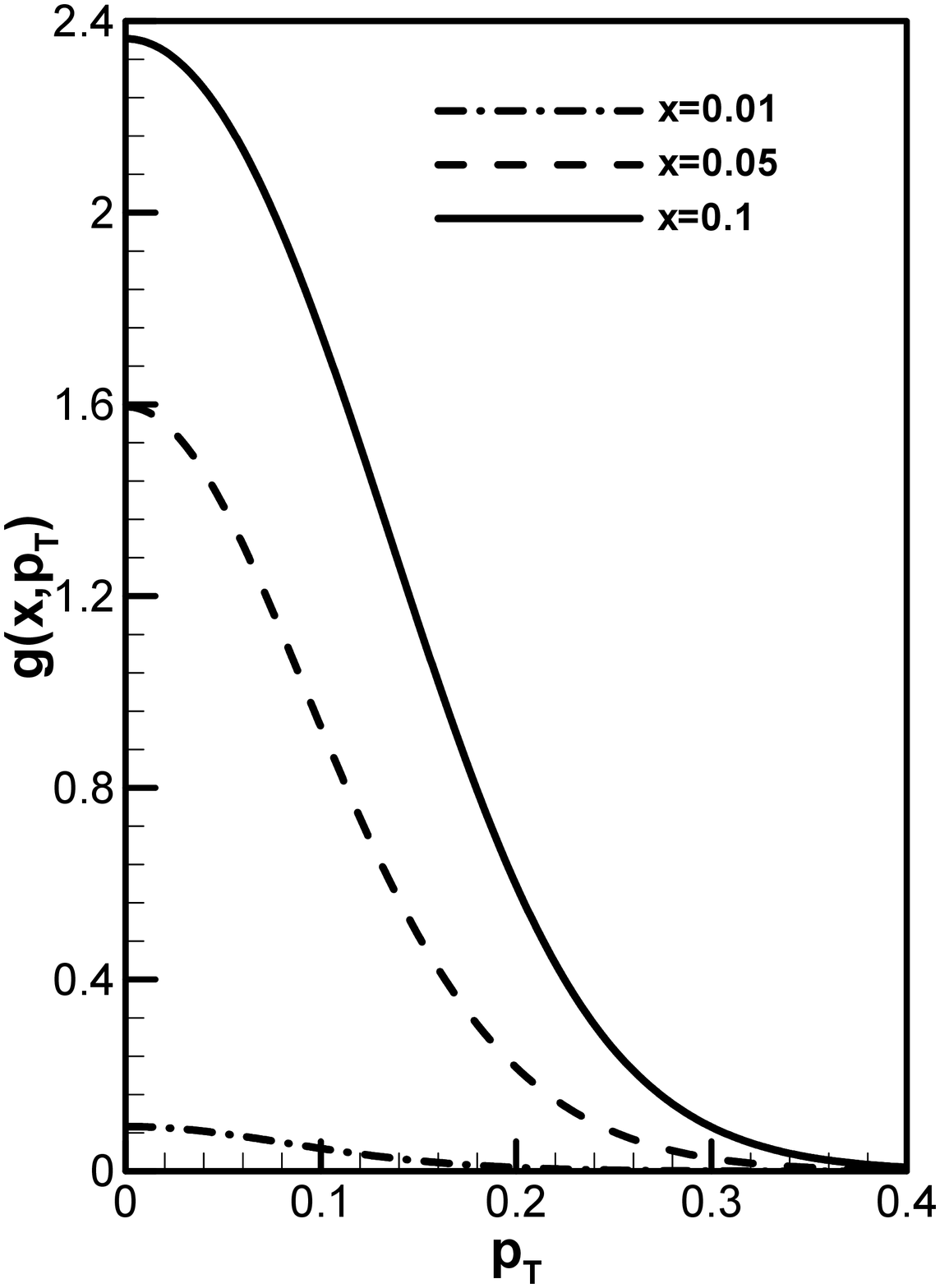}} &
       {\includegraphics[width=60mm,height=40mm]{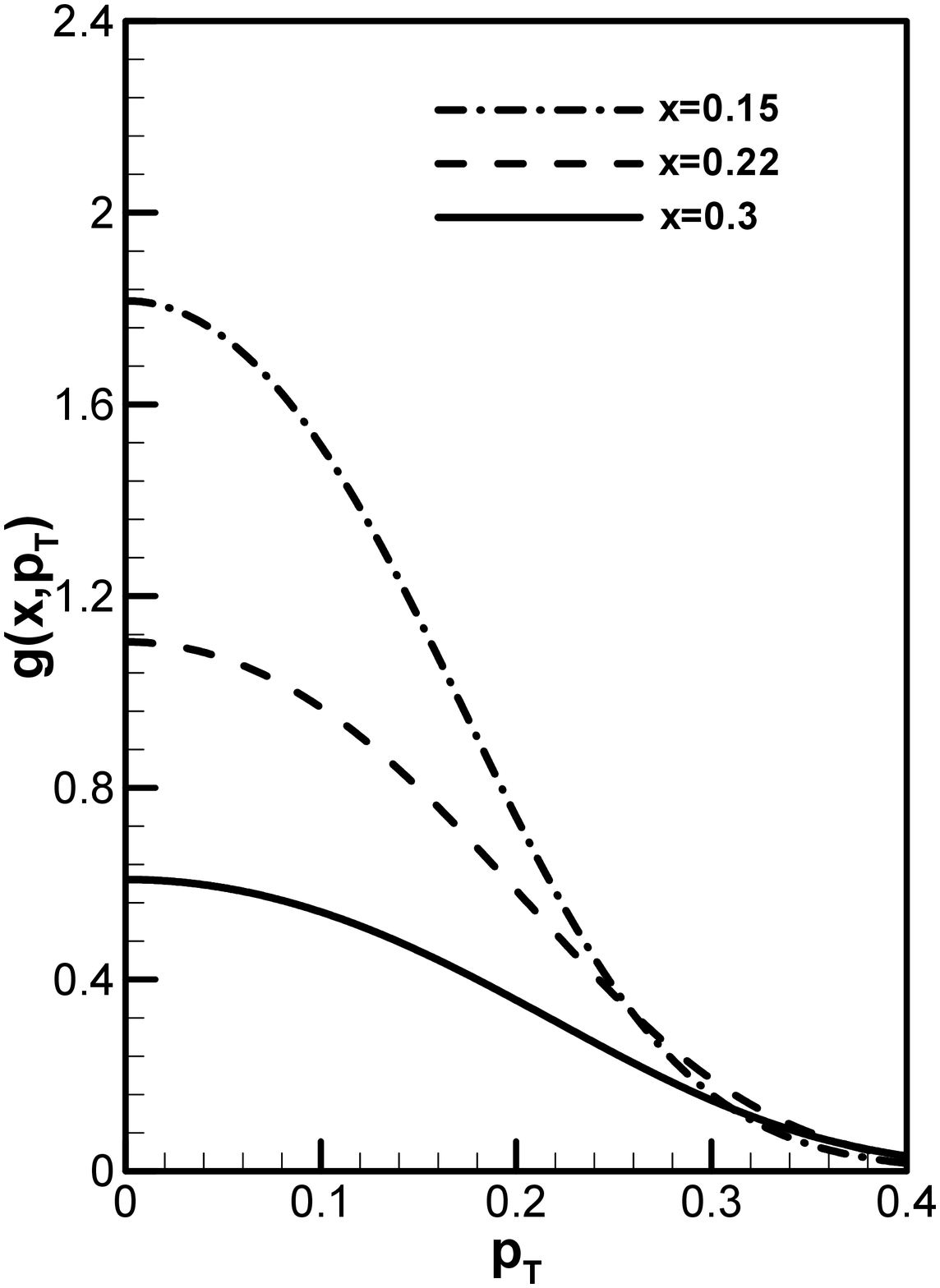}} \\
     \end{tabular}
\caption{ The TMD $s$ quark and gluon distributions with respect to $p_T$ at different values of $x$. \label{fig:6}}
      \end{center}
\end{figure*}
According to above discussion, the unpolarized TMD quark distributions are given as following:
\begin{eqnarray}
u(x,p_{T})&=&Z_{u}u_{0}(x,p_{T})+P_{u\pi^{-}/d}\otimes d_{0}\nonumber\\
&&+\frac{1}{2}P_{u\pi^{0}/u}\otimes u_{0}+P_{{ug}/{u}}\otimes u_{0},\label{u(x)}
\end{eqnarray}
\begin{eqnarray}
d(x,p_{T})&=&Z_{d}d_{0}(x,p_{T})+P_{d\pi^{+}/u}\otimes
u_{0}\nonumber\\
&&+ \frac{1}{2}P_{d\pi^{0}/d}\otimes
d_{0}+P_{{dg}/{d}}\otimes d_{0},\label{d(x)}
\end{eqnarray}
\begin{equation}
\hspace{-0.5 cm}s\left( x,p_{T} \right) = {P_{{{s{k^ + }}
\mathord{\left/
 {\vphantom {{s{k^ + }} u}} \right.
 \kern-\nulldelimiterspace} u}}} \otimes {u_0} + {P_{{{s{k^0}} \mathord{\left/
 {\vphantom {{s{k^0}} d}} \right.
 \kern-\nulldelimiterspace} d}}} \otimes {d_0} .~\label{sdis}
\end{equation}
$Z_u$ and $Z_d$ are the renormalization constants of the $u$ and $d$ bare quark distributions.\\
Finally, the TMD gluon distribution function is written as:
\begin{equation}
g(x,p_{T})=P_{{ug}/{u}}\otimes u_{0}+P_{{dg}/{d}}\otimes d_{0}.\label{g(x)}
\end{equation}
Now we are able to calculate the TMD quark and gluon densities using above relations.
\section{Results}
We calculate the unpolarized TMD quark and gluon distribution functions inside the proton using the effective chiral quark model at $Q^2=0.35~GeV^2$. To this end we first compute the TMD splitting functions and also bare quark densities which were discussed in previous section.

In Fig.\ref{fig:3}, we have displayed the three dimensional representation of the  TMD splitting functions $P_{u\pi^{-}/d}$, $P_{sk^{0}/d}$ and $P_{{ug}/{u}}$ with respect to $y$ and $p_{T}$.

We have depicted the TMD $u$, $d$ ad $s$ quark and also gluon distributions with respect to $x$ at three values of $p_{T}$ ($p_{T}=0.1, 0.2$ and $0.3~GeV$) in Fig.\ref{fig:4}.\\
It is shown that, as have been expected, by increasing the $p_{T}$ value, the TMD densities falloff down. In fact, the probability of finding the partons at larger values of $p_{T}$ is less. This behaviour of the TMD distributions is also seen in Refs.\cite{HA,AVE,AVE1,ZAVADA}.

In Fig.\ref{fig:5} and \ref{fig:6} we have plotted the TMD quark and gluon densities with respect to $p_{T}$ at different values of $x$. It is found that these $p_{T}$ distributions have the forms very close to the Gaussian distributions \cite{ZAVADA} and also the width of these densities is $x$ dependent \cite{ZAVADA}.\\
It should be pointed out that we have displayed each of $p_T$ densities in two set of $x$ values. In the first set which contain the small values of $x$, the TMD distributions grow by increasing the $x$ value in spite of the behaviour of the second set in which the $p_T$ densities decrease by increasing the amount of the $x$ variable.

Our results are arising out completely from a theoretical framework. As can be seen in Figs.\ref{fig:4}-\ref{fig:6}, these results yield us an appropriate behaviour in comparison with the results of Refs.\cite{HA,AVE,AVE1,ZAVADA}. Furthermore, we have also calculated the TMD function for strange($s$) quark and gluon distributions, using our theoretical model. Based on the current quark models, our investigations indicate that so far computations of the TMD distributions for gluon and $s$-quark have not been done.
\section{Conclusion}
A good understanding of partron densities can be obtained by studying  the deep-inelastic lepton nucleon scattering. These studies provide us how the momenta of parton densities are distributed parallel to the nucleon momenta.  To go   beyond the one dimensional consideration of  quark and gluon substructure, we need  to investigate the  transverse momentum dependent of parton densities. This can be done by taking into account the transverse momenta of produced hadrons via processes, for instance, semi-inclusive DIS or dileptons resulted from the Drell-Yan process.

In this article we used for the first time the modified $\chi QM$ to achieve the transverse momentum dependence of parton densities in unpolarized case. For this propose we used the Sullivan processes and suggested the TMD splitting functions which are being used in the $\chi QM$. Another key gradient in our calculations is to obtain the TMD bare quarks. This would be possible if we convert properly the measure of related integration to the light cone coordinate. What we got for the TMD parton densities are representing acceptable behaviour with respect to the variation of transverse momentum and the $x$-Bjorken variable. Extension the calculations to the polarized case, using the modified $\chi QM$ is possible which we hope to report them in future.

%----------------------------------------------------------------------------------------


\begin{thebibliography}{99}
\bibitem{SMC} SMC Collaboration, A. Bravar {\it{et al.}}, Nucl. Phys. A {\bf 666}, 314 (2000).
\bibitem{HER} HERMES Collaboration, A. Airapetian {\it{et al.}}, Phys. Rev. Lett. {\bf 94}, 012002 (2005).
\bibitem{COM} COMPASS Collaboration, V. Y. Alexakhin {\it{et al.}}, Phys. Rev. Lett. {\bf 94}, 202002 (2005).
\bibitem{JCDS} J. C. Collins and D. E. Soper, Nucl. Phys. B {\bf 194}, 445 (1982).
\bibitem{MB} M. Barkardt, Phys. Rev. D {\bf 62}, 071503(R) (2000).
\bibitem{RP} J. P. Ralston and B. Pire, Phys. Rev. D {\bf 66}, 111501 (2002).
\bibitem{MD} M. Diehl, Eur. Phys. J. C {\bf 25}, 223 (2002).
\bibitem{BJY} A. V. Belitsky, X. Ji and F. Yuan, Phys. Rev. D {\bf 69}, 074014 (2004).
\bibitem{myj} A. Mirjalili  {\it{et al.}}, J.~Phys.~G:~Nucl.~Part.~Phys. {\bf {37}}, 105003 (2010).
\bibitem{nymf} H. Nematollahi, M. M. Yazdanpanah and A. Mirjalili, J.~Phys.~G:~Nucl.~Part.~Phys. {\bf {39}}, 045009 (2012).
\bibitem{pasquini} B. Pasquini and S. Boffi, Phys. Rev. D {\bf 73}, 094001 (2006).
\bibitem{Suli} J. D. Sullivan, Phys. Rev. D {\bf 5}, 1732 (1972).
\bibitem{EHQ} E. J. Eichten, I. Hinchliffe and C. Quigg, Phys. Rev. D {\bf 45}, 2269 (1992).
\bibitem{weber} H. J. Weber, Int. J. Mod. Phys. A {\bf 14}, 3005 (1999).
\bibitem{weber1} H. J. Weber, Few. Body. Syst. {\bf 26}, 135 (1999).
\bibitem{swnpa} K. Suzuki and W. Weise, Nucl.~Phys.~A {\bf {634}}, 141 (1998).
\bibitem{MA2005} Y. Ding, R-G Xu and B-Q Ma, Phys. Rev. D {\bf{71}}, 094014 (2005).
\bibitem{MA2011} H. Song, X. Zhang and B-Q. Ma, Eur.~Phys.~J.~C {\bf {71}}, 1542 (2011).
\bibitem{phdt} K. Pumsa-ard, Chiral Dynamics of Baryons in the Perturbative Chiral Quark Models, Ph.D Thesis, Tuebingen (Germany) (2006); A. Faessler, Th. Gutsche, V. E. Lyubovitskij and K. Pumsa-ard, AIP Conference Proceedings 43 (2007).
\bibitem{ydm} M. M. Yazdanpanah, A. Mirjalili and M. Dehghanzadeh, Eur.~Phys.~J.~C {\bf {72}}, 2220 (2012).
\bibitem{MA2009} Y. Zhang, L. Shao and B-Q. Ma, Phys. Lett. B {\bf {671}}, 30 (2009).
\bibitem{HA} H. Avakin {\it{et al.}}, Phys. Rev. D {\bf {81}}, 074035 (2010).
\bibitem{AVE} A. V. Efremov {\it{et al.}}, arXiv:0912.3380 [hep-ph].
\bibitem{AVE1} A. V. Efremov {\it{et al.}}, Phys. Rev. D {\bf {83}}, 054025 (2011).
\bibitem{ZAVADA} P. Zavada, Phys. Rev. D {\bf {83}}, 014022 (2011).
\end{thebibliography}
\end{document}